\documentclass[usenatbib]{mn2e}
\usepackage{natbibmnfix,graphicx,times}

\newcommand{\Mpcden}{\mbox{ Mpc$^{-3}$}}
\newcommand{\Mpc}{\mbox{ Mpc}}

\newcommand{\Msun}{\mbox{ M$_\odot$}}

\newcommand{\hunits}{\mbox{ km s$^{-1}$ Mpc$^{-1}$}}
\newcommand{\bq}{\begin{equation}}
\newcommand{\eq}{\end{equation}}
\newcommand{\bqa}{\begin{eqnarray}}
\newcommand{\eqa}{\end{eqnarray}}

\newcommand{\mmin}{m_{\rm min}}
\newcommand{\bdgal}{\bar{\delta}_{\rm gal}}
\newcommand{\dL}{\delta^L}
\newcommand{\deriv}{{\rm d}}
\def\VEV#1{\left\langle #1\right\rangle}

\newcommand{\apj}{ApJ}
\newcommand{\apjl}{ApJ}
\newcommand{\apjs}{ApJS}

\newcommand{\aj}{AJ}
\newcommand{\mnras}{MNRAS}

\title[Cosmological galaxy voids]{The evidence of absence: galaxy voids in the excursion
set formalism}

\author[Furlanetto \& Piran]{Steven R. Furlanetto$^1$\thanks{Email: sfurlane@tapir.caltech.edu} \& Tsvi Piran$^{1,2}$ \\
$^1$Division of Physics, Mathematics, \& Astronomy; California Institute of Technology; Mail Code 130-33; Pasadena, CA 91125 \\
$^2$Racah Institute for Physics, Hebrew University, Jerusalem 91904, Israel.}

\begin{document}

\maketitle

\begin{abstract}
We present an analytic model for the sizes of voids in the galaxy
distribution.  Peebles and others have recently emphasized the
possibility that the observed characteristics of voids may point to a
problem in galaxy formation models, but testing these claims has been
difficult without any clear predictions for their properties.  In
order to address such questions, we build a model to describe the
distribution of galaxy underdensities.  Our model is based on the
``excursion set formalism,'' the same technique used to predict the
dark matter halo mass function.  We find that, because of bias, galaxy
voids are typically significantly larger than dark matter voids and
should fill most of the universe.  We show that voids selected from
catalogs of luminous galaxies should be larger than those selected
from faint galaxies: the characteristic radii range from $\sim 5$--$10
\, h^{-1} \Mpc$ for galaxies with absolute $r$-band magnitudes $M_r -
5 \log h < -16$ to $-20$.  These are reasonably close to, though
somewhat smaller than, the observed sizes.  The discrepancy may result
from the void selection algorithm or from their internal structure.
We also compute the halo populations inside voids.  We expect small
haloes ($M \la 10^{11} \Msun$) to be up to a factor of two less
underdense than the haloes of normal galaxies.  Within large voids, the
mass function is nearly independent of the size of the underdensity,
but finite-size effects play a significant role in small voids ($\la 7
\, h^{-1} \Mpc$).
\end{abstract}


\section{Introduction}
\label{intro}

One of the key predictions of any cosmological structure formation
model is the distribution of matter on large scales.  We now know that
the cold dark matter (CDM) paradigm can account for many of the
observed characteristics of the galaxy distribution.  In this picture,
the universe contained tiny (gaussian) density fluctuations at the
time the cosmic microwave background was last scattered.  Bound
structures assembled themselves through gravitational instability
around these perturbations in the relatively recent past.  In the past
two decades, with the advent of high-resolution numerical simulations,
it has become possible to follow this picture through the formation of
massive galaxies and clusters.  The CDM model accurately describes the
abundance and clustering of collapsed objects from dwarf galaxies to
rich galaxy clusters over a wide range of redshifts (although the details of galaxy formation itself
remain somewhat mysterious) as well as the distribution of neutral gas
in the intergalactic medium.

These systems all correspond to density peaks in the matter
distribution (with the exception of the lowest column density
Ly$\alpha$ forest absorbers).  For a variety of reasons, the other end
of the density distribution -- underdense voids -- has received
considerably less attention, despite their long observational history
\citep{gregory78,kirshner81} and their place as the most visually
striking features of the galaxy distribution.  This is largely because
voids subtend enormous volumes and so require large surveys to garner
representative samples.  Although voids have been found in every
redshift survey
\citep{delapparent86,vogeley94,elad97-iras,elad00,muller00,hoyle02},
the first statistically significant sample came only with the 2dF
redshift survey \citep{hoyle04}.  Now, with the DEEP2 redshift survey
and the Sloan Digital Sky Survey, it is even possible to constrain the
evolution of voids over the redshift interval $z \sim 1$--$0$
\citep{conroy05}.

\citet{hoyle04} presented the most complete search for voids to
date. They found that voids with characteristic radii $R \approx
15 \, h^{-1} \Mpc$ fill $\sim 35\%$ of the universe.  But their
search illustrates a second difficulty in studying voids: how to
define and identify them precisely (and meaningfully).
\citet{elad97} proposed the \emph{Voidfinder} algorithm based on
separating the observed galaxies into ``void'' and ``wall''
populations and building voids around gaps in the wall population
(see also \citealt{hoyle02}). While clearly defined for any given
observational sample, the results can nevertheless be difficult to
interpret in relation to the underlying physical quantities of
interest.  For example, the distribution of observed sizes depends
on the galaxy sample (intrinsically brighter galaxies yield larger
voids) as well as the search algorithm (\citealt{hoyle04} restrict
their search to radii greater than $10 \, h^{-1} \Mpc$, for
example).  Smaller voids are difficult to pinpoint because of
confusion with random fluctuations in the galaxy distribution.  We
will nevertheless follow this approach and define a void to be any
coherent region where the galaxy density falls below some
threshold.  Note that this differs from many other studies (e.g., \citealt{einasto89,gottlober03,sheth04})
that require a void to be \emph{completely} empty of galaxies.
These voids, which are much smaller than the voids we consider
here, do not correspond to the voids detected by eye in the
galaxy distribution.

The void phenomenon is also relatively difficult to study
theoretically.  The simplest model is a spherical tophat
underdensity. The early evolution of such a system is
well-described by spherical expansion (the analog of the
well-known spherical collapse model; \citealt{peebles80}).
Underdensities expand in comoving units, gradually deepening,
until they reach ``shell-crossing,'' when the center is evacuated
and individual mass shells cross paths.  At this stage, the 
initial spherical expansion model breaks down.  Later,  the
voids continue to expand relatively slowly in a self-similar
fashion \citep{fillmore84,suto84,bertschinger85}.  Unfortunately,
these dark matter models are idealizations in that real voids are
observed only through the galaxy distribution.  Because galaxies
are biased relative to the dark matter, we must ask what kind of
physical systems observed voids actually represent.  They are in
fact nearly empty of galaxies -- but does that require
shell-crossing, or can they be at an earlier evolutionary stage?

Another difficulty is that the large size of voids restricts the
usefulness of numerical simulations.  Early efforts focused on
understanding the dynamics of individual voids, which did not require
particularly high resolution \citep{dubinski93,vandeweygaert93}.  But
placing voids in their proper cosmological context demands both large
volumes and high mass resolution -- the latter because we must resolve
the galaxies from whose absence we identify voids (e.g.,
\citealt{goldberg04}).  Only recently have $N$-body simulations of the
required dynamic range become practical.  This has allowed the first
systematic studies of the structure of large voids \citep{gottlober03}
as well as of simulated voids in the dark matter distribution
\citep{colberg05} and in the galaxy distribution
\citep{mathis02,benson03}.  Nevertheless, there are still no full
hydrodynamic simulations of the void phenomenon: galaxy properties are
currently determined through semi-analytic models
\citep{mathis02,benson03}.

There are a number of reasons to study the void phenomenon.  Early
attention focused on using the observed voids to constrain
cosmological parameters.  \citet{blumenthal92} and \citet{piran93}
argued that the scale of typical voids would depend on the matter
power spectrum, just as the abundance of clusters does.  It was
quickly realized that the void distribution seemed to have much more
large-scale power than collapsed objects -- most obviously, although
voids with sizes $\ga 15 h^{-1} \Mpc$ are not uncommon, collapsed
objects do not reach the same mass scales.  This is puzzling because
an underdense region reaches shell-crossing only after the equivalent
overdensity would virialize.  \citet{friedmann01} pointed out that a
solution might lie in a proper treatment of the galaxies used to
define voids: galaxy bias could lower the required dark matter
underdensity, allowing voids to be larger for a given amount of CDM
power.  They argued that the observed population of large voids
preferred a $\Lambda$CDM model.

Recently, interest in voids has focused on their role in galaxy
formation.  \citet{peebles01} argued that (in the CDM model) voids should be populated by
small dark matter haloes but that the observed voids
appear to lack faint galaxies as well as bright ones.  Is this a
fundamental problem for the CDM paradigm, or does it indicate that
galaxy formation proceeds differently in voids?  One popular
explanation is that photoheating during reionization may have suppressed the
formation of dwarf galaxies in low-density environments
\citep{tully02,barkana04-fluc,cen05}, helping to clear faint galaxies
from the voids.  To study these problems, recent attention has focused
on the variation of the galaxy luminosity function with the
large-scale environment.  \citet{croton05} and \citet{hoyle05} (see
also \citealt{goldberg05}) found that the characteristic galaxy
luminosity and the galaxy density both decrease significantly in
low-density environments but that the faint-end slope remains nearly
constant.  The latter may be surprising given the expected variation
in the halo mass function with environment.

Before we can answer any of these questions, however, we require a
proper understanding of voids in CDM models.  \citet{mathis02} and
\citet{benson03} made important strides forward with their studies of
voids in $N$-body simulations.  They claimed that current
semi-analytic galaxy formation models predict void properties similar
to those observed and argued that the conflict pointed out by
\citet{peebles01} was illusory.  But such studies are still limited by
their dynamic range and by the (necessarily complex) galaxy formation
prescriptions imposed on the dark matter haloes.

The goal of this paper is to produce a straightforward analytic model
of the void distribution and of galaxy populations within voids.  Such
a model will sharpen our understanding of voids in a CDM model and
generate a baseline prediction with which we can contrast their
observed properties.  We will build on \citet{sheth04}, the most
compelling theoretical model of voids to date.  They used the excursion
set formalism, which reproduces the abundance of collapsed haloes
extremely well, to predict the distribution of void sizes.  However,
they required voids to reach shell-crossing and defined their
properties in terms of the dark-matter underdensity.  As a result,
they predicted characteristic void radii much smaller than those
observed.  Our main goal is to modify their approach so as to describe
voids in the \emph{galaxy} distribution.  Along the way, we will also
be able to predict the halo populations within voids and quantify
the claimed discrepancy with observational results.

The remainder of this paper is organized as follows.  In \S \ref{exp},
we examine the nonlinear evolution of the void density.  Then, in \S
\ref{defn}, we show how to compute the linearized underdensity of
voids with a specified galaxy underdensity, and we briefly discuss the
expected galaxy populations inside voids.  In \S \ref{abundance}, we
show how to compute the cosmological abundance of large \emph{galaxy}
voids and compare our results to observations.  Finally, we conclude
in \S \ref{disc}.

In our calculations, we assume a cosmology with $\Omega_m=0.3$,
$\Omega_\Lambda=0.7$, $\Omega_b=0.046$, $H=100 h \hunits$ (with
$h=0.7$), $n=1$, and $\sigma_8=0.9$, consistent with the most recent
measurements \citep{spergel03}.

\section{The Gravitational Expansion of Voids}
\label{exp}

Because the principal quantity of interest is the physical volume of
voids, we first describe how they
expand beyond their initial comoving size.  We will follow the method
of \citet{friedmann01}.  We consider the expansion of a tophat density
perturbation that begins with a (negative) density perturbation $\delta_i$
inside a physical radius $R_i$ at an initial time $t_i$ (corresponding
to $z_i$).  If we assume zero peculiar velocity perturbation and
consider only the epoch before shell-crossing (so that the mass inside
of each shell is conserved), conservation of energy yields the
equation of motion \citep{peebles80}
\begin{eqnarray}
\dot{R}^2(t) & = & H_0^2 [ - \Omega_0 (1+ z_i)^3 R_i^2 \delta_i  \nonumber \\ 
& & + \Omega_0  (1+z_i)^3 (1+\delta_i) R_i^3/R(t)
  + \Omega_\Lambda R(t)^2 ],
\label{eq:exp}
\end{eqnarray}
where the first term on the right hand side is (twice) the initial
total energy of the shell, the second term is (twice) the
gravitational potential energy at time $t$, and the last term is
(twice) the excess energy from the cosmological constant.  We solve
this equation for the physical radius of the shell as a function of
time or redshift.  The solution is such that the void expands in
comoving units once nonlinear effects set in.  We define
\begin{equation}
\eta \equiv \frac{R (1 + z)}{R_i (1+z_i)}
\label{eq:etadefn}
\end{equation}
to be the ratio of the comoving size at redshift $z$ to its initial
comoving size.  The real fractional underdensity due to gravitational
expansion is thus $1 + \delta = \eta^{-3}$.  Note that this is
independent of $R_i$.

Equation (\ref{eq:exp}) includes nonlinear expansion, but most of the
following will be phrased in terms of the equivalent linear density
$\dL$ extrapolated to the present day.  (For clarity, we will denote
all linear-extrapolated densities with a superscript ``L" in the
following.)  In order to transform to physical densities and scales,
we require $\eta(\dL)$ or equivalently $\delta(\dL)$.  We therefore
also compute the density as if linear theory were always accurate.
Any perturbation can be divided into growing and decaying modes; in
linear theory the growing mode (which is of course the only component
surviving to the present day) obeys \citep{heath77}
\begin{equation}
\dL = \frac{3 \delta_i H(z)}{5} \int_z^\infty \deriv u \frac{u+1}{H^3(1/u-1)},
\label{eq:grow}
\end{equation}
where $H^2(z) = \Omega_0(1+z)^3 + \Omega_\Lambda$ in a flat universe.
The constant $3/5$ comes from our choice of initial conditions: a
perturbation with zero peculiar velocity has $60\%$ of its amplitude
in the growing mode and the remainder in the decaying mode
\citep{peebles80}.

\begin{figure}
\begin{center}
\resizebox{8cm}{!}{\includegraphics{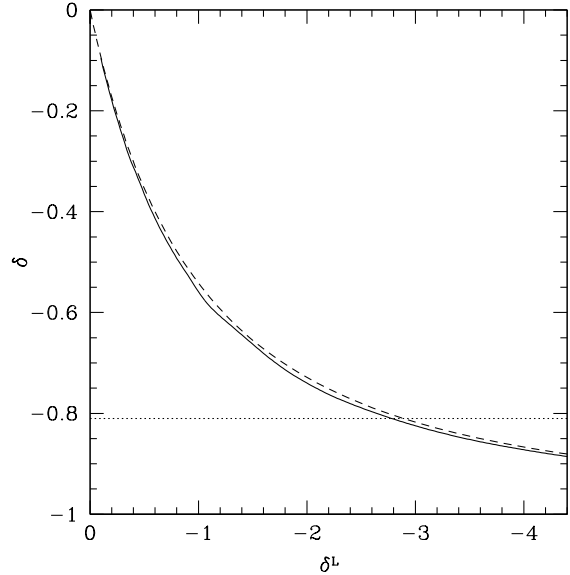}}\\%
\end{center}
\caption{Physical dark matter underdensity $\delta$ as a function of
  the linearized underdensity $\dL$ (solid line).  Note that our model
  actually breaks down for $\delta<-0.81$, when shell-crossing occurs
  (marked by the horizontal dotted line).  The dashed line shows the fitting function of \citet{mo96}.}
\label{fig:dL}
\end{figure}

We show the relation between the linearized and physical
underdensities with the solid line in Figure~\ref{fig:dL}.  At first, while linear
evolution is accurate, $\delta \approx \delta^L$, but then the
physical underdensity flattens out and even a small deepening of the
void corresponds to a large increase in $|\delta^L|$.  The dashed line in Figure~\ref{fig:dL} shows a fitting function for $\delta(\dL)$ from eq. (18) of \citet{mo96}.  Although those authors were interested in $\delta > 0$, the formula is accurate to a few per cent in the underdense regime as well.  For reference,
shell-crossing occurs at $\dL=-2.8$, corresponding to a physical
underdensity $\delta=-0.81$.  Beyond this point, our model for $\eta$
breaks down because the enclosed mass no longer remains constant.  The
expansion then proceeds self-similarly, with the comoving radius $R
\propto a^{1/3}$ in an Einstein-de Sitter universe
\citep{fillmore84,suto84,bertschinger85}, significantly slower than
the expansion before shell-crossing.  As a result, the void volume
fraction will be dominated by regions that have not yet (or have just)
reached shell-crossing \citep{blumenthal92}: deeper perturbations are
rarer in the CDM model and can be neglected because they expand so
slowly.  We therefore do not consider the later evolution in any more
detail.

For completeness, we also note that Figure~\ref{fig:dL} is nearly
independent of the input cosmological parameters; differences in the
cosmology manifest themselves in the growth rate of $\delta^L$ rather
than in the function $\delta(\delta^L)$ \citep{friedmann01}.

\section{Defining Voids}
\label{defn}

It may now seem straightforward to compute the void distribution: one
simply chooses a physical density threshold $\delta$, transforms to
$\dL$, and uses the excursion set formalism to determine the mass
distribution of objects with the specified linearized underdensity
\citep{sheth04}.  While such a procedure is perfectly well-defined, it
has one crucial shortcoming: in observed samples, we do not define
voids in terms of their dark matter density but in terms of their
\emph{galaxy} density.  Thus we actually want to relate the physical
galaxy underdensity $\bdgal$ within some region to $\dL$.  That is the
task of this section.

\subsection{The galaxy mass function}
\label{galfcn}

\begin{table*}
\begin{minipage}{130mm}
\caption{Mass Thresholds and Survey Depths}
\label{tab:numden}
\begin{tabular}{ccccc}
\hline
$M_r - 5 \log h$ & $M_{b_J} - 5 \log h$ & $n \, (10^{-3} \, h^3 \Mpcden)$ & $\mmin$ (PS;
$10^{11} \Msun$) & $\mmin$ (ST; $10^{11} \Msun$) \\
\hline
-16 & -16.3 & 58.6 & 1.0 & 0.73 \\ 
-18 & -17.8 & 27.9 & 2.5 & 1.7 \\ 
-19 & -18.7 & 15.2 & 5.2 & 3.4 \\ 
-20 & -19.5 & 5.89 & 15 & 10 \\ 
-21 & -20.4 & 1.11 & 90 & 60 \\
\hline
\end{tabular}
\medskip
\\
Number densities are computed from the \citet{blanton03} $r$-band and
\citet{croton05} $b_J$-band luminosity functions.  Columns 4 and 5
give the minimum halo mass required to match these densities, with the
\citet{kravtsov04} $\VEV{N(m_h)}$, for the Press-Schechter and
Sheth-Tormen mass functions, respectively.
\end{minipage}
\end{table*}

We will use the excursion set formalism to derive the dark matter halo
mass function \citep{bond91}.  In the simplest such model, a halo of
mass $m_h$ forms whenever the smoothed density field first exceeds a
linear-extrapolated density $\dL_c$ that is fixed by the physics of
spherical halo collapse ($\dL_c=1.69$ for objects collapsing at the
present day; \citealt{peebles80}).  To solve such a problem, we
consider diffusion in the $(\dL,\,\sigma^2)$ plane, where $\sigma^2$
is the variance of the density field smoothed on scale $m_h$.
Trajectories begin with $\dL=0$ at $\sigma^2=0$ (i.e., they must have
the mean density on infinitely large scales).  They then diffuse away
from the origin as density modes are added on smaller scales.  The
problem is simply to compute the distribution of $\sigma^2$ (or
equivalently $m_h$) at which these trajectories cross the absorbing
barrier $\dL_c$ for the first time.  From this first-crossing
distribution, the mean comoving number density of haloes with masses $m_h
\pm \deriv m_h/2$ is \citep{press74,bond91}
\begin{equation}
n_h(m_h) = \sqrt{ \frac{2}{\pi} } \, \frac{\bar{\rho}}{m_h^2} \,
\left| \frac{\deriv \ln \sigma}{\deriv \ln m_h} \right| \,
\frac{\dL_c}{\sigma} \, \exp \left[ - \frac{(\dL_c)^2}{2 \sigma^2}
\right],
\label{eq:nmh}
\end{equation}
where $\bar{\rho}$ is the mean matter density.  In a region with
linearized underdensity $\dL_v$ and mass $M_v$ [with $M_v = (4 \pi/3)
\, \bar{\rho} \, (R_v/\eta)^3$ for a void], the comoving number
density of haloes is \citep{bond91,lacey93}
\begin{eqnarray}
n_h(m_h|\dL_v,M_v) & = & \sqrt{ \frac{2}{\pi} } \,
\frac{\bar{\rho}}{m_h^2} \, \left| \frac{\deriv \ln \sigma}{\deriv \ln
m_h} \right| \,
\frac{\sigma^2(\dL_c-\dL_v)}{(\sigma^2-\sigma^2_v)^{3/2}} \nonumber \\ 
& & \times \exp
\left[ - \frac{(\dL_c - \dL_v)^2}{2 (\sigma^2 - \sigma^2_v)} \right],
\label{eq:nmcond}
\end{eqnarray}
where $\sigma_v \equiv \sigma(M_v)$ and of necessity $\sigma > \sigma_v$.  This conditional mass function
follows from an identical diffusion problem, but in this case the
trajectories begin at $(\dL_v,\sigma^2_v)$ rather than the origin.

Although equation (\ref{eq:nmh}) provides a reasonable match to halo
abundances in cosmological simulations, it is by no means perfect.
\citet{sheth99} and \citet{jenkins01} provide more accurate fits to
the simulation results.  We will nevertheless use the standard
Press-Schechter abundances in the following.  The main reason is that,
although the Sheth-Tormen mass function can also be motivated by a
diffusion problem with the absorbing barrier fixed by ellipsoidal
collapse, there is no analytic form for the conditional mass function
because the barrier changes its effective shape with a shift of the
origin.  \citet{sheth02} showed that the conditional mass function
could be approximated through a Taylor series, but in that case the
corresponding unconditional mass function differs slightly from the
usual Sheth-Tormen form (which provides the best match to numerical simulations).  \citet{zhang05} showed how to derive the
mass function corresponding to such barriers using a simple numerical
scheme.  But for our purposes the Press-Schechter form suffices.  The
most important property is the density dependence.  \citet{barkana04-fluc} and
\citet{furl05-charsize} found that it is extremely
similar to that contained in equation (\ref{eq:nmcond}), at least at
high redshifts, and the former also showed that it fits simulations
fairly well.

Equations (\ref{eq:nmh}) and (\ref{eq:nmcond}) give the dark matter
halo abundance, which is not the same as the galaxy abundance.  To
connect the two we require the mean number of galaxies per halo, $\VEV
{N(m_h)}$, the first moment of the halo occupation distribution.  For
simplicity, we will use the universal form of \citet{kravtsov04}.
They split the mean occupation number into two parts.  The first,
$\VEV {N_c(m_h)}$, represents the number of central galaxies in the
halo and is unity if $m_h>\mmin$ and zero otherwise.  We will normally
choose $\mmin$ by comparison to some observational detection
threshold; note therefore that it need not reflect the \emph{actual}
minimum mass of a galaxy.  The second part, $\VEV {N_s(m_h)}$,
describes the number of satellite galaxies above the same detection
limit.  \citet{kravtsov04} found that this has a simple form
\begin{equation}
\VEV {N_s(m_h)} = \left( \frac{m_h}{C \, \mmin} \right)^{\beta}.
\label{eq:ns}
\end{equation}
From fits to the subhalo distribution in $N$-body simulations, they
estimate that $C \approx 30$ and $\beta \approx 1$ at $z=0$ over a
broad range of halo masses.  We will use these fiducial values in most
of our calculations.  In order to estimate the importance of the halo
occupation distribution, we will also compare to a model with
$\VEV{N(m_h)}=1$.

The total comoving number density of (observable) galaxies within a
region is therefore
\begin{equation}
n_g^c(\mmin|\dL_v,M_v) = \int_{\mmin}^\infty \deriv m_h \,
\VEV{N(m_h)} \, n_h(m_h|\dL_v,M_v),
\label{eq:Nmh}
\end{equation}
where the superscript ``c'' denotes a cumulative quantity.  We wish to
compare to voids selected through galaxy surveys.  We will therefore
set $\mmin$ by comparison to the galaxy number densities in the Sloan
Digital Sky Survey\footnote{See http://www.sdss.org/.} and the 2dF
Galaxy Redshift Survey.\footnote{See
http://www.mso.anu.edu.au/2dFGRS/.}  Specifically, we match to the
number density of galaxies with $r$-band luminosity greater than a
specified value, using the luminosity function of \citet{blanton03},
and to the corresponding quantities from the $b_J$-band luminosity
function of \citet{croton05}.  The appropriate mass thresholds $\mmin$
for the Press-Schechter and Sheth-Tormen mass functions are shown in
Table \ref{tab:numden}.  We will also show some results for
$\mmin=10^{10} \Msun$ in order to mimic an exceptionally deep survey.
Note that normalizing in this way -- to the cumulative number density
of galaxies -- tends to wash out differences between mass functions,
insulating us from much of the uncertainty in using the
Press-Schechter mass function.

\subsection{The linearized underdensity of voids}
\label{dv}

The total (observed) galaxy underdensity in a void with physical size
$R_v$ is
\begin{equation}
1 + \bdgal(\mmin,\dL_v,R_v) = \frac{n_g^c(\mmin|\dL_v,M_v)}{\eta^3 \,
n_g^c(\mmin)}.
\label{eq:bdgal}
\end{equation}
Given $\mmin$, we solve this equation to find the linearized
underdensity required to produce a void of size $R_v$ and mean
observed galaxy underdensity $\bdgal$.

\begin{figure}
\begin{center}
\resizebox{8cm}{!}{\includegraphics{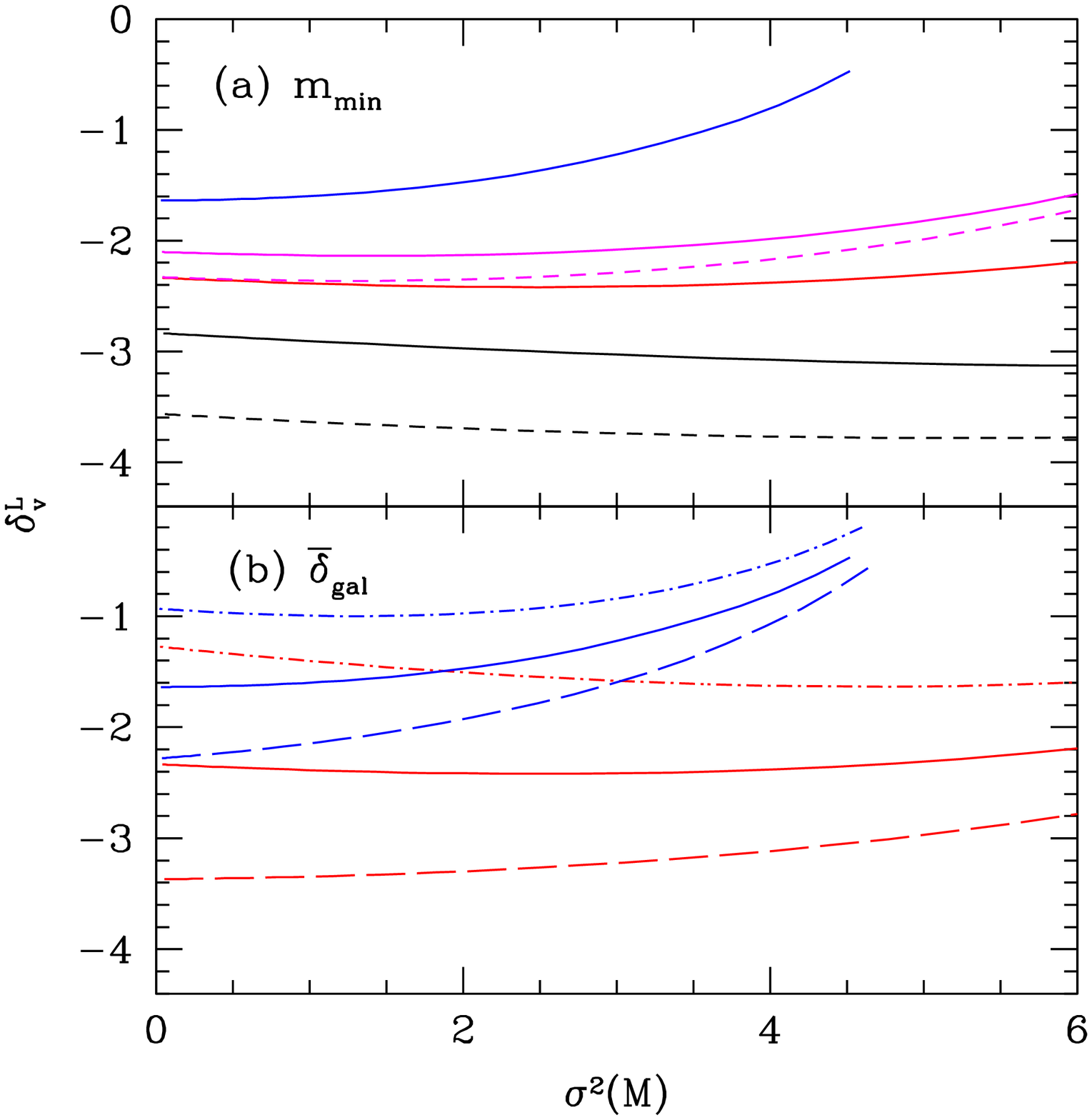}}\\%
\end{center}
\caption{Linearized underdensity $\dL_v$ required to produce a mean
(physical) galaxy underdensity $\bdgal$.  \emph{(a)}: All curves
assume $\bdgal=-0.8$.  The solid curves use our fiducial model, with
$M_r<-20,\,-18,\,-16$, and $m_h>10^{10} \Msun$, from top to bottom.
The dashed curves assume one galaxy per dark matter halo; we show
$M_r<-18$ and $m_h>10^{10} \Msun$.  \emph{(b)}: The dot-dashed, solid,
and dashed curves assume $\bdgal=-0.6,\,-0.8$, and $-0.9$,
respectively.  We show results for $M_r<-20,\,-16$ (top and bottom
sets of curves).}
\label{fig:barriers}
\end{figure}

Figure~\ref{fig:barriers} shows the results of this inversion.  The
solid curves in panel \emph{(a)} give the required underdensity for
several different survey depths: $M_r<-20,\,-18,\,-16$, and
$m_h>10^{10} \Msun$, from top to bottom.  (Here and throughout, we
will suppress the ``$-5 \log h$" in measured magnitudes for the sake
of brevity.)  We take $\bdgal=-0.8$ as a fiducial value.  We plot
$\dL_v$ as a function of mass variance $\sigma^2$.  For reference,
$\sigma^2=0.4,\,1.9$, and $4$ correspond to comoving sizes $R \approx
15,\,5$, and $2.5 h^{-1} \Mpc$, respectively.  Note that this is
\emph{not} the physical size of the void (which is larger by a factor
$\eta$ and varies with $\dL_v$).  We halt each curve at $M_v = 2
\mmin$.  The required underdensity becomes more extreme in deeper
surveys; this is because underdense regions are much more deficient in
massive galaxies than small haloes in the extended Press-Schechter
formalism.  Haloes with masses $m_h \la M_v$ cannot collapse because
too few density modes are available between $(M_v,\,m_h)$ to pass the
collapse threshold $\dL_c$, but if $\sigma^2 \gg \sigma_v^2$ (for
small haloes) the loss of large-scale power becomes unimportant.  In
other words, the bias increases with galaxy mass, so the large
galaxies are more sensitive to the underlying density field \citep{efstathiou88, cole89, mo96}.  This is
why $\dL_v$ curves upward as $\sigma^2$ increases, especially in the
$M_r<-20$ case.  There is also a slight trend for $\dL_v$ to increase
as $\sigma \rightarrow 0$; this is because the shape of the mass
function changes in underdense regions, with more mass in small haloes.
As the void size approaches infinity, slightly more of the mass shifts
into galaxy-mass haloes.

The two dashed curves in Figure~\ref{fig:barriers}\emph{a} show
$\dL_v$ if we assume $\VEV{N(m_h)}=1$ for $M_r<-18$ and $m_h>10^{10}
\Msun$.  This decreases the threshold, because it increases the weight
of the smallest haloes (which are least sensitive to the density).  The
value of $\dL_v$ becomes more sensitive to the halo occupation number
for deeper surveys; a larger minimum mass pushes the survey to the
steep part of the mass function, where the extra satellites in
extremely large (but rare) haloes make only a small difference.  Note
as well that in this case the $m_h>10^{10} \Msun$ threshold lies
significantly below shell-crossing, where our model breaks down.

Figure~\ref{fig:barriers}\emph{b} varies $\bdgal$ for $M_r<-20$ and
$M_r<-16$ (top and bottom sets of curves).  We let
$\bdgal=-0.6,\,-0.8$, and $-0.9$ for the dot-dashed, solid, and dashed
curves.  This obviously has a strong effect on $\dL_v$, because the
massive galaxies are so sensitive to the underlying dark matter
density.  Note that the differences are largest at small $\sigma^2$
(or large physical scales), because at smaller scales the finite size
of the void limits the maximum halo mass.  This is also why the
differences are larger for the deeper survey.

\subsection{Haloes in voids}
\label{gal}

Now that we have defined voids, we can look more closely at their
resident halo populations.  Figure~\ref{fig:barriers} clearly shows
that the intrinsic properties of voids depend sensitively on the mass
limit of the survey used to detect them: deeper surveys require larger
dark matter underdensities and hence will presumably find smaller
voids.  But, in a given void, will smaller haloes be more abundant
relative to the (brighter) galaxy population used to define the void?
To answer this question, we let
\begin{equation}
1 + \delta_{h}(m) = \frac{n_h(m|\dL_v,M_v)}{\eta^3 \, n_h(m)}.
\label{eq:dgal}
\end{equation}
We emphasize that $\delta_{h}$ is the halo underdensity at mass $m$,
\emph{not} the galaxy underdensity.  The analogous galaxy underdensity
at a particular mass is more difficult to calculate, because it
requires the distribution of galaxy masses within each halo instead of
simply the number of galaxies (i.e., how many satellite galaxies of
mass $m$ are contained in any galaxy group).  Such models require
considerably more machinery than is appropriate for our simple
analytic model.  We will therefore content ourselves with computing
the halo distribution, which will illustrate the most important
features anyway.  We will comment on its relation to the galaxy
population below.

\begin{figure}
\begin{center}
\resizebox{8cm}{!}{\includegraphics{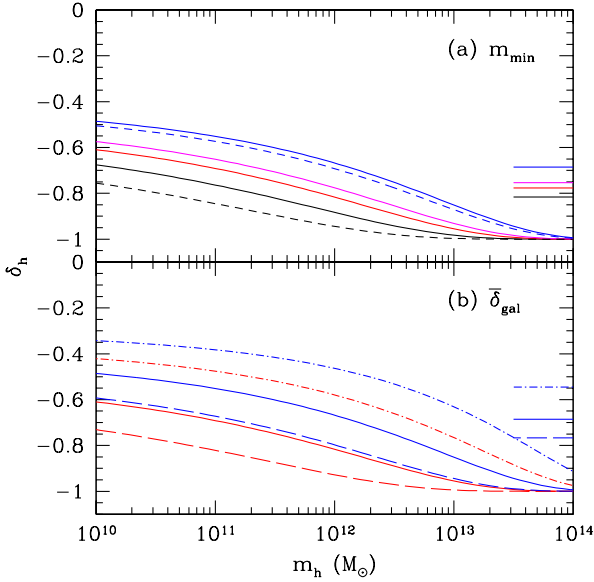}}\\%
\end{center}
\caption{Underdensity of \emph{haloes} within voids of physical size
$R_v=21 \, h^{-1} \Mpc$.  \emph{(a)}: Assumes $\bdgal=-0.8$.  The
solid curves use our fiducial model, with $M_r<-20,\,-18,\,-16$, and
$m_h>10^{10} \Msun$, from top to bottom.  The horizontal bars on the
right show $\delta$ in each of these voids (same ordering from top to
bottom).  The dashed curves assume one galaxy per dark matter halo; we
show $M_r<-20$ and $m_h>10^{10} \Msun$.  \emph{(b)}: The dot-dashed,
solid, and dashed curves assume $\bdgal=-0.6,\,-0.8$, and $-0.9$,
respectively.  We show results for $M_r<-20,\,-16$ (top and bottom
sets of curves).  The horizontal bars show the corresponding $\delta$
for the $M_r<-20$ cases.}
\label{fig:nmh}
\end{figure}

Figure~\ref{fig:nmh} shows predictions for the halo underdensity in
voids with $R_v=21 \, h^{-1} \Mpc$.  The parameters in the two panels
are identical to those in Figure~\ref{fig:barriers}.  In panel
\emph{(a)}, we see that small haloes become less underdense if void
selection is performed with larger galaxies.  Within any given void,
we also see that small haloes are less underdense than large haloes.
This is, of course, because larger haloes are more biased than small
ones.  The disparity can be quite large for the $M_r<-20$ sample, with
small haloes only half as underdense as those with $m_h \approx 10^{13}
\Msun$.  The dashed curves show that these conclusions are
qualitatively independent of $\VEV{N(m_h)}$.  In panel \emph{(b)}, we
see that varying $\bdgal$ also modifies the fractional overdensity.
However, its effects on the shape of the curve are relatively modest
except at the most massive end.

The horizontal bars on the right axis of Figure~\ref{fig:nmh} show the
physical dark matter underdensity within each void.  Interestingly, it
takes a relatively narrow range of values in panel \emph{(a)},
clustered around $\delta \approx -(0.7$--$0.8$), with $|\delta|$
decreasing as $\mmin$ increases.  The halo populations (especially at
large masses) spread over a much broader range in $\delta_h$.  This
is essentially a consequence of the shape of $\delta(\delta^L)$ shown
in Figure~\ref{fig:dL}: a narrow range in physical underdensities
spans a large range of $\delta^L$, which is the quantity relevant to
our model for halo abundances.  In general, $\delta \sim \delta_h$ for
haloes selected near $\mmin$.  The density of haloes with $m \ll \mmin$
is closer to the mean because their bias is less than unity.

\begin{figure}
\begin{center}
\resizebox{8cm}{!}{\includegraphics{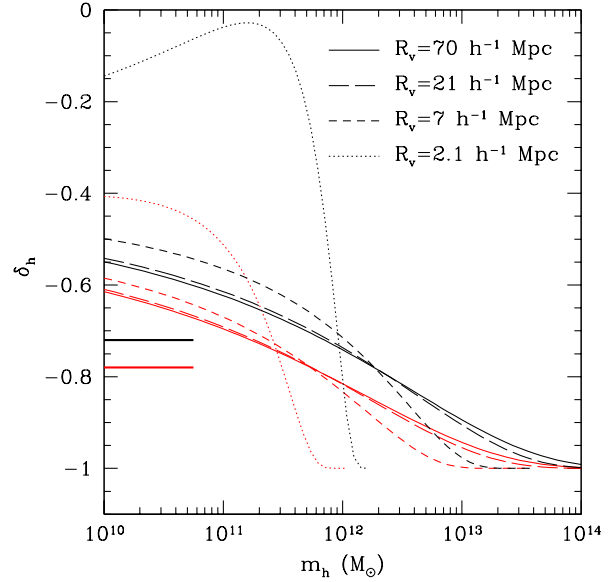}}\\%
\end{center}
\caption{Underdensity of \emph{haloes} within voids of $\bdgal=-0.8$
and several different radii.  The upper thick and lower thin sets of curves
assume $M_r < -19$ and $-16$, respectively.  The horizontal bars at
left show $\delta$ for $R_v=7 \, h^{-1} \Mpc$ voids.}
\label{fig:nmhr}
\end{figure}

Our model also predicts that $\delta_h$ at a fixed mass will vary with
the size of the void, because $\dL_v$ is also a function of scale.  We
show the implications in Figure~\ref{fig:nmhr}; in this case we
consider surveys with $M_r<-19$ and $M_r<-16$ (top and bottom sets,
respectively).  For large voids, $R_v \ga 7 \, h^{-1} \Mpc$, we see
only a slight steepening of the curves, which occurs because the
finite mass of the void limits the maximum halo mass.  Thus, galaxy
populations in voids with $R_v \ga 7 \, h^{-1} \Mpc$ should be nearly
independent of the void radius.  This is a consequence of the flatness
of $\dL_v$ at small $\sigma^2$ in Figure~\ref{fig:barriers}, which
occurs because so little power exists on such large physical scales.
However, when $R_v \sim 2.1 \, h^{-1} \Mpc$, the shapes change
dramatically because the finite size of the void strongly limits the
haloes inside.  A comoving volume with radius $R=2.1 \, h^{-1} \Mpc$
contains $\sim 10^{13} \Msun$, which is only $\sim 20$--$100$ times
larger than the mass limits of these surveys.  Thus the region
requires only a modest underdensity to decrease the abundance of
massive galaxies.  Such ``voids'' therefore sit near the mean density
and smaller haloes (for which finite size effects can be neglected)
will have nearly their mean density.  This simplified model suggests
that many properties of voids will depend on their spatial extents, at
least if $R_v \la 7 h^{-1} \Mpc$.

One consistent theme of this section is that large haloes are
relatively less abundant than small haloes within voids.  We can
compare this expectation to observations, which suggest that the
mean number density $\bar{n}_{\rm gal}$ and the characteristic
luminosity $L_\star$ decrease significantly in voids but that the
faint-end slope of the luminosity function remains nearly constant
\citep{croton05,hoyle05}.  Unfortunately, the comparison is not
trivial.  These studies measured the galaxy environment through
the mean density within spheres of size $\sim 8 \, h^{-1} \Mpc$
(generally by comparing to the density field of a volume-limited
sample of galaxies).  This is just large enough that finite size
effects should not be dramatic, but it is still in a regime in
which Poisson fluctuations in the galaxy distribution cannot be
neglected.  As a result, some of the ``low-density'' regions may
in fact be at the mean density, which would wash out the void
effect.  An ideal test would focus on those galaxies identified to
be within large voids (for example using the
\emph{Voidfinder} algorithm). Moreover, we must also bear in mind
that $\delta_h \neq \delta_{\rm gal}$. In particular, many small
galaxies will be satellites inside large haloes; such a population
will be efficiently suppressed in just the same way as bright
central galaxies.  Thus we expect the galaxy luminosity function
in voids to be flatter than the halo mass function.

With these caveats, our model naturally explains the decrease in
$\bar{n}_{\rm gal}$ and $L_\star$, especially because these regions
are relatively small and finite size effects help suppress galaxies
with $m \ga 10^{12} \Msun$.  Note that this probably explains the
discrepancy between the observed $L_\star$ and that predicted by
\citet{cooray05}, who let $R_v \rightarrow \infty$ rather than
including a cutoff at $R_v \sim 8 h^{-1} \Mpc$ which would have
quenched the formation of massive galaxies.  On the other hand, our
models robustly predict an excess (by about a factor of two) between
the abundance of small and large haloes in voids.  If there were a
one-to-one correspondence between halo mass and galaxy luminosity, and
if the luminosity function could be approximated by a power law $n
\propto L^{-\alpha}$ over this interval, we would expect $\alpha$ to
be $\sim 0.15$ larger in voids than in mean density environments.  The
suppression of satellite galaxies would reduce the disparity.  Even
so, our result is within the error bars of \citet{croton05}, though
their best-fit values indicate no significant change.  Thus, there is
currently only weak evidence, at best, for non-standard galaxy
formation within voids.  CDM models robustly predict an excess of
small haloes inside of voids, but it is not particularly strong and
requires care to interpret.

\section{Void Abundances}
\label{abundance}

\subsection{The excursion set abundance}
\label{excurs}

\citet{sheth04} used the excursion set formalism to estimate the
number density of voids.  They defined a void as a region with dark
matter underdensity $\dL_v$.  At first sight, the void abundance seems
to follow from equation (\ref{eq:nmh}), but with $\dL_c \rightarrow
\dL_v$: this is the same problem as the halo abundance, except that
the absorbing barrier is negative rather than positive.  However,
\citet{sheth04} pointed out one crucial difference, which they called
the ``void-in-cloud" problem.  Consider a small region with $\dL <
\dL_v$ that is contained in a larger region with $\dL > \dL_c$.
Equation (\ref{eq:nmh}) would have assigned such a point to a small
void; however, physically we know that this ``void" lies inside of a
collapsed object and hence has been crushed out of existence.  (This
problem does not occur for $n_h$ because haloes are allowed to collapse
inside voids.)

Thus, the appropriate diffusion problem has \emph{two} barriers: we
wish to compute the first-crossing distribution for the void barrier
$\dL_v$ while excluding trajectories that have already crossed a
barrier with $\dL_p>0$ that describes void-crushing.  The simplest
assumption is to take both these barriers to be independent of
$\sigma^2$; in that case \citet{sheth04} showed that the mass function
is (see also the alternate derivation in the Appendix)
\begin{eqnarray}
n_v(m_v) & = & \frac{\bar{\rho}}{m_v^2} \, \left| \frac{ \deriv \ln
\sigma}{\deriv \ln m_v} \right| \sum_{n=1}^\infty \left\{ \frac{n^2 \pi^2
D^2}{(\dL_v/\sigma)^2} \, \frac{\sin(n \pi D)}{n \pi} \right. \nonumber \\
& & \times \left. \exp \left[ -
\frac{n^2 \pi^2 D^2}{2 (\dL_v/\sigma)^2} \right] \right\},
\label{eq:nvm}
\end{eqnarray}
where
\begin{equation}
D \equiv \frac{| \dL_v |}{\dL_p + | \dL_v |}
\label{eq:Ddefn}
\end{equation}
describes the relative importance of void-crushing; a fraction $(1-D)$
of all matter lies inside voids in this model.  Note that $\dL_v$
and $\dL_p$ implicitly depend on redshift, because we evaluate $\sigma$
at the present day.

The \citet{sheth04} model contains one crucial assumption -- that
voids can be defined at a constant $\dL_v$, independent of scale --
and two free parameters.  In what follows we will usually set the
void-crushing parameter $\dL_p=1.06$, which is the linearized
overdensity at turnaround according to the spherical collapse model.
This is reasonable because it marks the point at which larger scale
overdensities will begin to collapse around their resident voids, but
we will also let $\dL_p$ take larger values for illustrative purposes.
The second free parameter is $\dL_v$; \citet{sheth04} argued that
$\dL_v=-2.8$ (shell-crossing) is an appropriate choice, because that
marks the maximum point of efficient expansion.

However, we have already seen that the dark matter underdensity is
generally significantly deeper than the galaxy underdensity.  Instead,
if one wishes to compare to voids observed in redshift surveys,
$\dL_v$ should be chosen from the $\bdgal$ and $\mmin$ appropriate to
a given void search.  The relevant absorbing barriers are precisely
the curves shown in Figure~\ref{fig:barriers}: only in voids selected
from exceptionally deep galaxy surveys do we expect shell-crossing to
have occurred.  Interestingly, the barriers are also relatively flat,
at least for reasonably deep surveys and on large scales.  We will
therefore take $\dL_v=$\,constant and use equation (\ref{eq:nvm}) to
compute the number density of voids.  We evaluate $\dL_v$ at $R_v=21
\, h^{-1} \Mpc$ in the following; we examine how $n_v(m_v)$ varies
with the barrier shape in \S \ref{assump} below.

Figure~\ref{fig:nvoid} shows the void abundance for several different
survey depths.  Panel \emph{a} shows the fraction of volume filled by
voids of a given radius; note that we use the physical void radius
here, including the extra gravitational expansion.  Panel \emph{b}
shows the volume filling fraction of voids,
\begin{equation}
F_{\rm vol}(>R) = \eta^3 \int_{m}^\infty \deriv m_v \,
\frac{m_v}{\bar{\rho}} \, n_v(m_v),
\label{eq:fvol}
\end{equation}
where the lower integration limit is the mass corresponding to a void
of physical radius $R$; the prefactor $\eta^3$ accounts for the extra
gravitational expansion.  We halt each curve when the void mass is less
than twice $\mmin$.

\begin{figure}
\begin{center}
\resizebox{8cm}{!}{\includegraphics{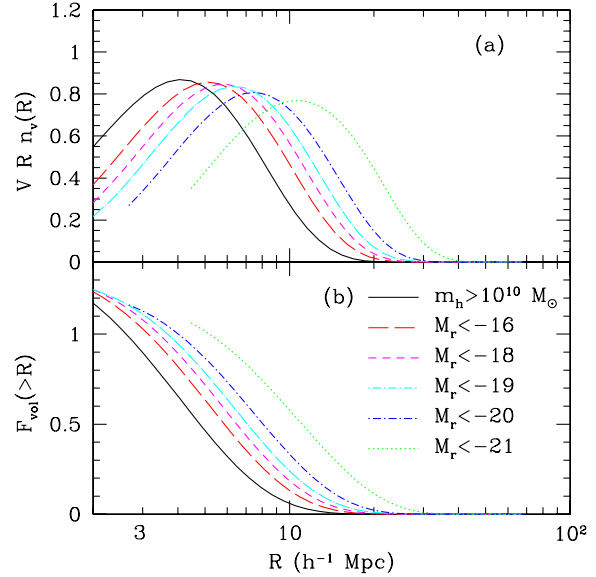}}\\%
\end{center}
\caption{Abundance of voids in the excursion set formalism.  All
curves follow our fiducial model; they vary the minimum detectable
galaxy mass as shown in the legend.  Panels \emph{a} and \emph{b} show
differential and cumulative filling factors of voids, respectively.}
\label{fig:nvoid}
\end{figure}

Figure~\ref{fig:nvoid} makes clear several key characteristics of the
void size distribution.  First, $n_v$ has both large and small mass
cutoffs.  The former occurs where $\sigma \approx \dL_v$, above which
voids are exponentially suppressed because of the lack of large-scale
power (the same mechanism that provides the large-scale cutoff in the
halo mass function).  The small mass cutoff occurs because of the
void-crushing threshold $\dL_p$.  The resulting characteristic scale
$R_c$ is quite sensitive to the galaxies used to select voids: it
ranges from $R_c \approx 5 h^{-1} \Mpc$ for $M_r<-16$ to $R_c \approx
10 h^{-1} \Mpc$ for $M_r<-21$, nearly an order of magnitude in volume.
The shape, on the other hand, does not change much across the
different selection thresholds.  Finally, we also see that $F_{\rm
vol}>1$: the voids apparently fill a volume larger than the universe.
This occurs because we have let each and every void expand to its full
size, even those surrounded by relatively weak overdensities just
short of turnaround.  These filling fractions should thus not be taken
too seriously; they suggest only that, when voids are selected from
galaxy surveys, we expect empty regions with $R \ga 3 h^{-1} \Mpc$ to
fill a large fraction of the universe.

The largest possible voids also provide an interesting benchmark for
comparison to surveys \citep{blumenthal92}.  To estimate $R_{\rm
max}$, we find where $n_v = (c/H_0)^{-3}$; i.e., the size for which we
expect one void per Hubble volume.  This will likely overestimate the
maximum size in a real survey, because voids were smaller in the past
(see \S \ref{redshift} below).  Of course the maximum size will also
vary with the magnitude limit of the survey; for
$M_r<(-16,\,-19,\,-21)$ we find $R_{\rm max} \approx (28,\,32,\,45) \,
h^{-1} \Mpc$.  These are compatible with existing redshift surveys;
the largest voids in the 2dF survey have $R \sim 25 h^{-1} \Mpc$
\citep{hoyle04}.  We will make a more detailed comparison to
observations in \S \ref{obs} below.

Finally, we note some simple properties of $F_{\rm vol}$.  Ignoring
void-crushing, the void filling fraction becomes
\begin{equation}
F_{\rm vol}^{\rm nc}(>R) \approx \eta^3 \, {\rm erfc} \left[
\frac{\dL_v(z)}{\sqrt{2} \sigma(R)} \right].
\label{eq:fvapprox}
\end{equation}
This is the same expression found by \citet{dubinski93}, except that
they assumed $\dL_v$ corresponded to shell crossing and included an
extra factor of one-half (which occurred because they ignored those
trajectories that joined larger voids).  Void-crushing is unimportant
for voids larger than the characteristic size (see \S \ref{assump}
below), so equation (\ref{eq:fvapprox}) provides a reasonable estimate
for $F_{\rm vol}$ and shows explicitly how it depends on
$\dL_v$ (and hence on the survey characteristics).  \citet{dubinski93}
pointed out the interesting coincidence that $F_{\rm vol} \approx 1$
at the present day; this remains true in our model.  Only recently
have large voids in the faint galaxy population come to dominate the
geography of the universe.

\subsection{Void depth and abundance}
\label{depth}

\begin{figure}
\begin{center}
\resizebox{8cm}{!}{\includegraphics{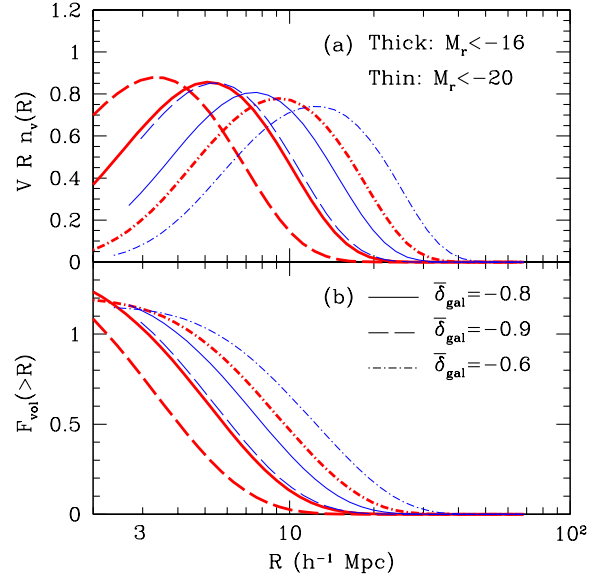}}\\%
\end{center}
\caption{As Fig.~\ref{fig:nvoid}, except we vary the mean galaxy
underdensity within the voids: $\bdgal=-0.8,\,-0.9$, and $-0.6$
(solid, dashed, and dot-dashed curves, respectively).  The thick and
thin sets of curves take $M_r<-16$ and $M_r<-20$. }
\label{fig:nvdgal}
\end{figure}

We have seen that one key quantity in defining voids is the mean
enclosed galaxy underdensity $\bdgal$.  Figure~\ref{fig:nvdgal} shows
how the void sizes vary with this criterion.  The curves correspond to
the same cases shown in Figure~\ref{fig:barriers}\emph{b}.  We see
that $\bdgal$ has an enormous effect on the characteristic void size
$R_c$: it changes by a factor of three over this range of galaxy
density (or a factor of nearly thirty in volume!).  Moreover, in panel
\emph{(b}), we see that large, deep voids ($\bdgal=-0.9$) only fill a
small fraction of space.

\subsection{Voids at higher redshifts}
\label{redshift}

\begin{figure}
\begin{center}
\resizebox{8cm}{!}{\includegraphics{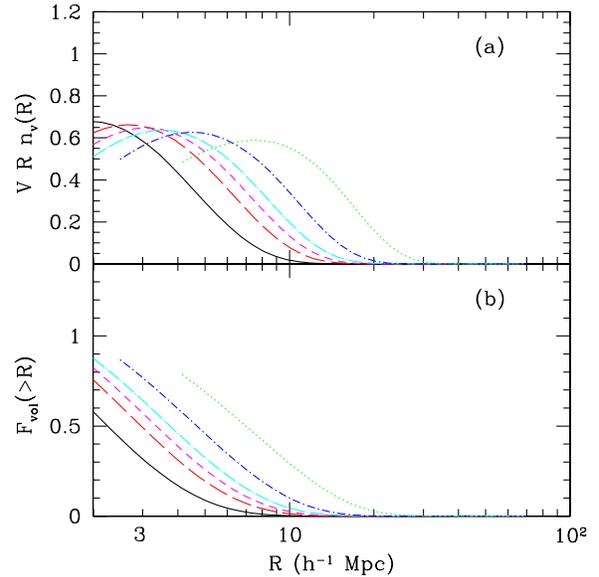}}\\%
\end{center}
\caption{As Fig.~\ref{fig:nvoid}, except for $z=1$.  The galaxies are
  selected with the same \emph{mass} thresholds as in
  Fig.~\ref{fig:nvoid} or Table~\ref{tab:numden} (with the solid curve
  having $\mmin>10^{10} \Msun$ and mass increasing to the right).
  Note that the $z=1$ number densities differ from the $z=0$ values.}
\label{fig:nvz}
\end{figure}

Just as with the halo mass function, the excursion set formalism also
allows us to study how $n_v(m)$ evolves with redshift.  We show the
predicted void distribution at $z=1$ in Figure~\ref{fig:nvz}.  In
order to facilitate comparison with the $z=0$ results, we hold $\mmin$
constant between $z=1$ and the present day (as in Table 1).  Of
course, because the galaxy luminosity function evolves over this
interval, the \emph{number density} of these haloes evolves over this
redshift interval.

The voids also evolve, with the characteristic radius increasing
significantly (by about a factor of two in each case) and the volume
filling fraction also increasing (again by a factor of $\sim 2$--$3$).
For example, only $\sim 10\%$ of the universe lies inside of voids
with $R>10 \, h^{-1} \Mpc$ selected from $m_h>1.5 \times 10^{12}
\Msun$ galaxies (as opposed to $\sim 33\%$ at $z=0$).  This implies
that voids continue to grow, albeit relatively slowly, even after
the cosmological constant dominates the energy density.  Our results
are in qualitative agreement with \citet{conroy05}, who find that
voids are both smaller and rarer at $z \sim 1$.  A precise comparison
is difficult because they use statistical techniques (specifically the
void probability function) rather than identifying individual voids.

\subsection{Model assumptions}
\label{assump}

\begin{figure}
\begin{center}
\resizebox{8cm}{!}{\includegraphics{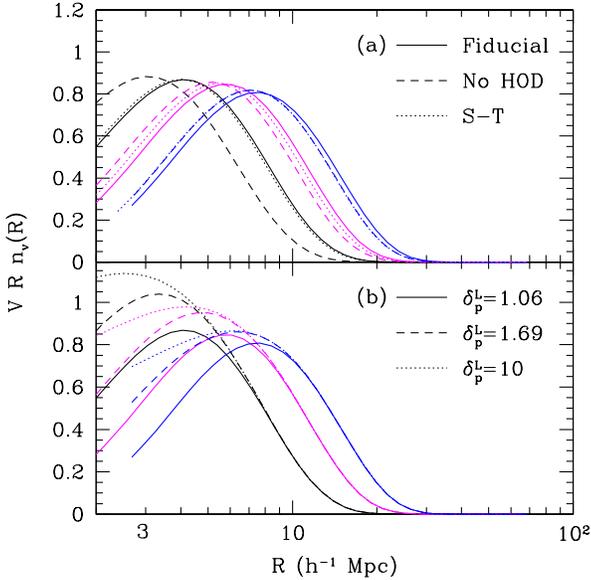}}\\%
\end{center}
\caption{Void abundance for several parameter choices.  All curves
take our fiducial model except as noted.  \emph{(a)}: Dashed curves
assume one galaxy per dark matter halo, while the dotted curves use
the Sheth-Tormen mass function \emph{(b)}: The solid, dashed, and
dotted curves set the ``void-crushing" parameter $\dL_p=1.06,\,1.69$,
and $10$, respectively.  In both panels, the three sets of curves
assume surveys with $\mmin=10^{10} \Msun$, $M_r<-18$, and $M_r<-20$,
from left to right.}
\label{fig:nvdparam}
\end{figure}

Unfortunately, this model for $n_v(m)$ contains a number of free
parameters and simplifications, so it is not nearly as well-specified
as the halo mass function $n_h(m)$.  Here we will examine how
sensitive our results are to these assumptions.  We begin with the
halo occupation number $\VEV{N(m_h)}$.  The solid and dashed curves in
Figure~\ref{fig:nvdparam}\emph{a} contrast the \citet{kravtsov04}
value (our fiducial model) and a model with one galaxy per halo.  We
have previously seen that $\VEV{N}$ only significantly affects $\dL_v$
if $\mmin$ is relatively small; as a result, $n_v(m)$ is only
substantially affected for the $\mmin>10^{10} \Msun$ survey.  In this
case the voids shrink by about $40\%$ in radius, with $R_c \sim 3 h^{-1}
\Mpc$.  Thus, if we selected voids from the \emph{halo} underdensity
(as, for example, may be possible by associating groups of galaxies
with single haloes), we would expect slightly smaller voids.  The
difference is minimal for bright galaxies because the mass function is
steep in that regime.

Figure~\ref{fig:nvdparam}\emph{a} also compares the void sizes
computed from the Press-Schechter mass function with those from the
Sheth-Tormen mass function.  To perform this comparison, we have
scaled the Sheth-Tormen mass function with density in the same way as
the Press-Schechter mass function \citep{barkana04-fluc}.  Thus the only
real difference is the mass threshold $\mmin$, although that difference can be
substantial.  We see that $n_v(m)$ remains nearly unchanged.
Thus while the void distribution may in detail depend on the exact
form of $n_h(m)$, it is not likely to be substantially affected.

Another unknown parameter in the model is $\dL_p$, which describes
void-crushing.  Figure~\ref{fig:nvdparam}\emph{b} shows its effects.
\citet{sheth04} argue that $\dL_p$ is unlikely to be smaller than
$1.06$, because voids should continue to increase in size until
turnaround.  We therefore consider $\dL_p=1.69$ (where voids are only
destroyed when they lie inside of collapsed haloes) and $\dL_p=10$
(where void crushing is negligible).  This parameter clearly does have
an effect on $n_v(m)$: ignoring void crushing dramatically increases
the number of small voids.  However, it has no effect on large scales,
because these voids cannot be inside of extremely dense regions anyway.
It also has little effect on the characteristic size $R_c$.  Thus,
while $\dL_p$ may affect the distribution of small voids, it is
unlikely to affect the much more striking population of large voids.

\begin{figure}
\begin{center}
\resizebox{8cm}{!}{\includegraphics{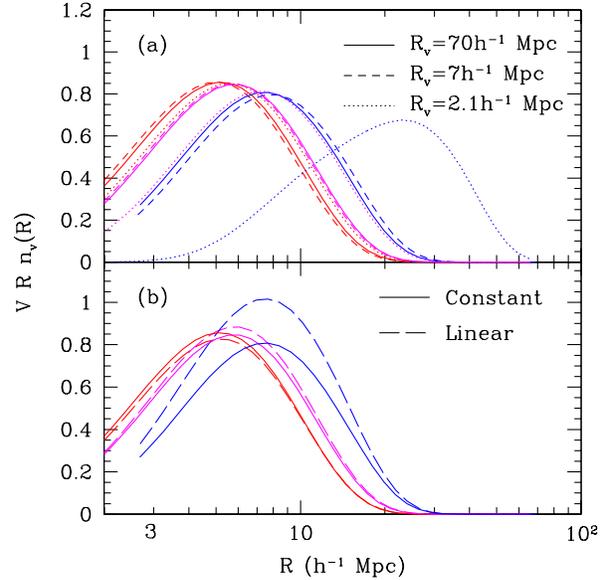}}\\%
\end{center}
\caption{Void abundance for different barrier prescriptions.  All
curves take our fiducial model except as noted.  \emph{(a)}: Solid,
dashed, and dotted curves set the (constant) excursion set barrier by
evaluating $\delta_v$ at $R_v=70,\,7$, and $2.1 h^{-1} \Mpc$,
respectively.  (The fiducial model uses $R_v=21 h^{-1} \Mpc$.)
\emph{(b)}: The dashed curves use a linear fit to the excursion set
barrier to compute the void abundance.  In both panels, the three sets
of curves assume surveys with $M_r<-16,\,-18$, and $-20$, from left to
right.}
\label{fig:nvdbarr}
\end{figure}

Perhaps the most important simplification we have made is to ignore
the scale dependence of $\dL_v$ so that we could use equation
(\ref{eq:nvm}).  Figure~\ref{fig:barriers} shows that the threshold
actually varies with the mass of the void.  In principle, we should
find the exact size distribution by following a procedure similar to
\citet{zhang05}, but with two absorbing barriers.  However, it is
relatively easy to see that in most cases the constant barrier
approximation suffices for our purposes, though more sophisticated
comparisons in the future may require a more exact solution.  One
straightforward test is to take $\dL_v$ to be constant but to evaluate
its amplitude at several different scales.  In
Figures~\ref{fig:nvoid}--\ref{fig:nvdparam}, we have used $\dL_v(R=21
h^{-1} \Mpc)$.  In Figure~\ref{fig:nvdbarr}\emph{a}, we show the
corresponding distributions for the threshold evaluated at $R=70,\,7$,
and $2.1 h^{-1} \Mpc$ (solid, dashed, and dotted curves).  We show
results for $M_r<-16,\,-18$, and $-20$ (left, middle, and right sets).
Clearly, for deep surveys $n_v(m)$ is nearly independent of the scale
at which we evaluate the barrier.  This is, of course, because the
corresponding thresholds in Figure~\ref{fig:barriers}\emph{a} are
nearly independent of $\sigma^2$.

However, if $M_r<-20$, the barrier rises rapidly at small scales.
From Figure~\ref{fig:nvdbarr}\emph{a}, the distribution is still
nearly independent of scale so long as $R_v \ga 7 h^{-1} \Mpc$, but
using $\dL_v(R=2.1 h^{-1} \Mpc)$ gives entirely different
results.  Of course, this choice is not self-consistent, because
the distribution is dominated by much larger voids.  To better
estimate the effects of this ``moving'' barrier, we note that it is
approximately a linear function in $\sigma^2$.  We would therefore
ideally like to solve for the mass function in the presence of a
constant barrier $\dL_p$ and a linearly increasing barrier $\dL_v +
\beta \sigma^2$.  Unfortunately, we were unable to find an analytic
solution for that distribution (though a straightforward numerical
solution may be possible by extending \citealt{zhang05}).  We
therefore instead turn to a model in which the void-crushing barrier
is also linear with the same slope as the void barrier.  In the
Appendix, we show that the appropriate mass function is
\begin{equation}
n_v^\ell(m_v) = n_v(m_v) \exp(-\beta \dL_v - \beta^2 \sigma^2/2).
\label{eq:nvmlin}
\end{equation}
(Note that this differs from the solution presented by
\citealt{sheth04}, which corresponds to a case with two barriers of
opposite slopes; see the Appendix for a more detailed discussion.)
Treating $\dL_p$ as a linear barrier is probably not a bad
approximation.  For one, the amplitude of $\dL_p$ does not strongly
affect $n_v(m)$, as shown by Figure~\ref{fig:nvdparam}\emph{b}.
Moreover, most trajectories that cross $\dL_v$ would have crossed the
void-crushing barrier at small $\sigma^2$, when $\beta \sigma^2 \ll
1$.  Finally, the appropriate choice of $\dL_p$ is somewhat arbitrary
anyway, so a linear model may actually be more accurate.

The similarity of $n_v^\ell(m_v)$ and $n_v(m)$ suggests that the
linear barrier will have only a small effect on the distribution.  In
the cases of interest, $\beta>0$, so the first exponential factor will
be larger than unity and more mass will lie inside of voids.  This is
intuitively obvious, because an increasing negative barrier is easier
to cross.  The second term affects the shape of the mass function, but
it is only important for large $\beta^2 \sigma^2$; in the regime we
study, this factor is always smaller than unity.
Figure~\ref{fig:nvdbarr}\emph{b} confirms these expectations.  For
$M_r<-20$, the shape remains nearly invariant (except at the smallest
scales) but more mass lies inside of voids.  Most importantly, the
characteristic scale $R_c$ is nearly identical in the two cases.  The
linear barrier makes no difference to fainter galaxy samples, because
the corresponding barrier is so nearly flat.

\subsection{Comparison to observations}
\label{obs}

We are now in a position to compare our model to the observed
distribution of voids.  \citet{hoyle04} compiled the most complete
sample to date from the 2dF Galaxy Redshift Survey.  They used the
\emph{Voidfinder} algorithm \citep{elad97,hoyle02} to identify
voids with $R>10 \, h^{-1} \Mpc$ in a volume-limited sample of
galaxies with $M_{b_J} \la -19$.\footnote{\citet{hoyle02} do not explicitly
state the absolute magnitude limit for their main sample, but
comparison of its size to their Table 2 yields this value.}  They
found that $\sim 35\%$ of the universe is contained inside voids
with $R>10 \, h^{-1} \Mpc$; the mean effective radius of their
sample was $R \approx 15 \, h^{-1} \Mpc$, with a maximum of $R
\approx 25 h^{-1} \Mpc$.  Their algorithm separates galaxies into
``wall'' and ``void'' populations; the latter can sit inside of
voids but the former are excluded.  They measured a mean galaxy
underdensity $\bdgal=-0.93$ within voids.

With this magnitude limit, the \citet{hoyle04} sample should find
voids somewhere between the two dot-dashed curves in
Figure~\ref{fig:nvoid} or near the thin curves in
Figure~\ref{fig:nvdgal}.  Given the measured $\bdgal=-0.93$, the most
naive comparison is to the $\bdgal=-0.9$ curve in the latter, which
has $F_{\rm vol}(R>10 \, h^{-1} \Mpc) \approx 16\%$ and $R_c \sim 5 \,
h^{-1} \Mpc$.  Both of these are considerably smaller than the
observed value, but it is unclear how significant the discrepancy is.
Unfortunately, identifying voids at the predicted characteristic scale
is difficult, because Poisson fluctuations in the galaxy distribution
are substantial on Mpc scales.  As a result, it is impossible to know
whether an observed galaxy underdensity with $R \sim 5 h^{-1} \Mpc$
corresponds to a true dark matter underdensity or a statistical
fluctuation.  Largely for this reason, \citet{hoyle04} restrict their
search to voids with $R>10 \, h^{-1} \Mpc$.  We therefore do not
necessarily consider the small $R_c$ predicted by our model to
conflict with observations; however, the discrepancy in $F_{\rm vol}$
for large voids is more worrying.  We do remark that a large
population of small voids have clear observational signatures,
increasing the number of $\sim 5 h^{-1} \Mpc$ empty or nearly empty
regions well above those expected in a random distribution.  This
would manifest itself in such statistical measures as the void
probability function (the probability that a region has zero galaxies;
\citealt{white79}) or the underdensity probability function (the
probability that a region has a given underdensity;
\citealt{vogeley89}), but such statistics are (at least for the
moment) beyond the capabilities of our model.

There are several more subtleties in a comparison to observations.
Most importantly, the \emph{Voidfinder} algorithm does not use a
density threshold, so the appropriate choice of $\bdgal$ is not
obvious.  The only ``void'' galaxies allowed in this approach are
those that are well-isolated.  But not all galaxies inside real
voids need be isolated: if some fraction of the wall galaxies are
actually contained inside regions with $\dL < \dL_v$, the true
voids would be less underdense than allowed by the algorithm.
Indeed, the $\bdgal=-0.8$ curves in Figure~\ref{fig:nvoid} predict
$F_{\rm vol}(R>10 \, h^{-1} \Mpc) \approx 25$--$33\%$ and $R_c
\sim 7$--$8 \, h^{-1} \Mpc$, much closer to the observed values.
There are solid physical reasons to expect confusion between
genuine void and wall galaxies.  Spherical underdensities tend to
evolve by evacuating their interiors, with mass accumulating in a
shell near the edge (see, e.g., \citealt{dubinski93}).
\citet{gottlober03} showed that massive haloes pile up at the edges
even more strongly.  Thus it is reasonable to expect that many of
the bright galaxies belonging to a large-scale underdensity lie
near its edges, where they are difficult to identify
unambiguously.  Moreover, real surveys operate in redshift space.
Because voids are expanding more rapidly than the Hubble flow,
their observed volumes will be larger than the physical volumes
used by our model.  \citet{goldberg04} show that the enhancement
can be $\ga 20\%$ near shell-crossing.

An equally important consideration is the geometrical method used
by \emph{Voidfinder} to identify voids.  It fills the gaps between
wall galaxies with spheres and then merges overlapping spheres
into discrete entities.  This procedure may join two neighboring
voids that our approach would consider distinct, especially if the
wall between them is weak or if we include Poisson fluctuations.
Galaxies on void walls tend to escape along the walls when
neighboring voids merge \citep{dubinski93}, making this picture
even more likely.  Finally, voids are not uniform; rather, they
are composed of a patchwork of (deeper) voids separated by weak
filaments \citep{gottlober03}.  In the excursion set picture, this
is the analog of the progenitor distribution of collapsed objects
and could in principle be computed through similar techniques
\citep{lacey93}, with the important difference that ``sub-voids''
can more easily retain their distinct identity as deeper
underdensities in the galaxy distribution.  The interaction of
this structure with void selection algorithms is not trivial.

Another clear prediction of our model is that voids become larger as
$\mmin$ increases.  \citet{hoyle04} addressed this question with
statistical techniques.  Specifically, they computed the void
probability function and underdensity probability function for
volume-limited samples with a range of magnitude thresholds (from
$M_{b_J} < -16$ to $M_{b_J}<-21$).  They found that both the fraction
of the universe filled by voids and their characteristic size
increase with the luminosity threshold, in qualitative agreement with
our results.  A more precise comparison requires an extension of our
model to these statistics.

In summary, although our model predicts a significant population of
large voids, they are still somewhat smaller than the observed
structures.  This could be a real discrepancy -- in which case voids
clearly require more sophisticated modeling -- or it could be a
combination of the contrasting selection techniqes, the internal void
structure, and redshift-space effects.  However, our qualitative
conclusions (large voids, increasing in size as the underlying galaxy
population brightens) do match the observed trends, a significant step
forward for analytic models.  Unfortunately, precise comparisons to
observations will still require numerical simulations.

\subsection{Comparison to simulations}
\label{sims}

Another useful comparison is to voids inside cosmological
simulations. In particular, \citet{benson03} used the
\emph{Voidfinder} algorithm to identify voids inside an $N$-body
simulation.  They used a semi-analytic model to place galaxies
within dark matter haloes and selected the voids based on the
predicted properties of the galaxies. They presented the void size
distributions for several different survey depths.  Our results
compare favorably with theirs.  Like us, they found larger voids
for brighter galaxies as well as larger voids for galaxies
compared to the dark matter (though below $\sim 10 \, h^{-1} \Mpc$
the two appear to converge; this may be related to the increasing
importance of Poisson fluctuations).  They also found that the
size distribution only peaks at $R \ga 10 \, \, h^{-1} \Mpc$ for
the brightest galaxies ($M_r<-21.5$ and $M_{b_J}<-20.5$ in their
models); for fainter galaxies their distributions continue rising
to $R \sim 7 \, h^{-1} \Mpc$ (below which their selection
technique becomes incomplete).  Finally, they find that voids
typically have $\delta \approx -0.8$ and $\bdgal \approx -0.9$,
with both quantities decreasing rapidly toward the center of the
void.  The density increases sharply near, but slightly beyond,
the nominal void radius. This provides a hint that the density
structure may play a role in the \emph{Voidfinder} selection,
making direct comparisons with our predictions somewhat more
difficult.  Nevertheless, the good qualitative, and even
reasonable quantitative, agreement of our model with these
simulations provides strong support for the major features of our
approach.

\citet{colberg05} recently examined voids in a series of large-volume
$N$-body simulations.  They found voids to be much smaller than our
predictions, with $90\%$ of the volume contained in voids with $R<2.5
\, h^{-1} \Mpc$.  However, they selected voids based on a dark-matter
density criterion $\delta_v<-0.8$ (corresponding to shell-crossing).
This is much more restrictive than our galaxy-density criterion.  The
best comparison is to the $m_h > 10^{10} \Msun$ survey in
Figure~\ref{fig:nvoid}, whose $\dL_v$ is close to shell-crossing (see
Fig.~\ref{fig:barriers}).  For this sample, we find $R_c \approx 4
h^{-1} \Mpc$, with much of the universe in smaller voids.  The
remaining discrepancy may result from the search algorithm.  They
began their void searches around local minima in the density field,
extending them until the enclosed density exceeded $\delta_v$.  Such a
prescription may tend to find ``sub-void'' progenitors rather than the
larger structures that correspond to the observed voids in the galaxy
distribution.

\section{Discussion}
\label{disc}

We have described a simple analytic model for voids in the galaxy
distribution.  Our model is based on \citet{sheth04}, who showed how
to apply the excursion set formalism to underdensities.  Its most important
parameter is $\dL_v$, the linearized underdensity of
a void.  Those authors originally set $\dL_v$ to be the density
corresponding to shell-crossing.  However, this condition produces
voids much smaller than the observed structures, whose masses
can exceed those of galaxy clusters \citep{piran93}.  Our major
contribution has been to show how to define $\dL_v$ through the
\emph{galaxy} underdensity.  Because galaxies are biased relative to
the dark matter, voids can be nearer to the mean density than the
observations naively indicate.  This significantly increases the
characteristic void size and the volume filling fraction.

Our model predicts voids with characteristic radii $R_c \sim 5$--$10
h^{-1} \Mpc$.  This is similar to, though somewhat smaller than, the
voids found in galaxy redshift surveys \citep{hoyle02}.  It is a
closer match to the void population in semi-analytic galaxy formation
models \citep{benson03}.  Because bright galaxies are more highly
biased, our model also predicts that voids selected from shallow
surveys should be characteristically larger, in agreement with
observations \citep{hoyle04}.  However, we also predict larger galaxy
densities inside voids than observed.  The significance of these
discrepancies is not clear: we argued in \S \ref{obs} that the
algorithms used to identify voids in surveys, redshift-space
distortions, and the internal structure of voids are all important in
detailed comparisons to the observations.  The last of these is
especially crucial, because simulations show that galaxies inside
voids tend to congregate near their edges \citep{gottlober03}, where
they are difficult to separate from genuine ``wall'' galaxies.  Two
other subtleties may also affect the comparison.  The first is the
nonlinear evolution of substructure within the void; we have used
linear theory to describe the halo population inside voids, which
likely breaks down in some regimes because power can be transferred
between scales.  The second is Poisson variation in the galaxy number
counts \citep{sheth-lemson99,casas02}.  This effect makes direct
detection of small voids impossible, forcing existing surveys to
search only for voids with $R \ga 10 h^{-1} \Mpc$.

Thus it is difficult to make a precise comparison between our simple
model and observed voids.  However, the qualitiative agreement is
reasonable.  Our analytic approach is the first to contain voids with
characteristic radii $\sim 10 \, h^{-1} \Mpc$ and hence to be in even
qualitative agreement with the observations.  Our model shows that the
enormous extent of observed voids is not particularly surprising and
should not be perceived as a ``crisis'' for the CDM model.

By using the excursion set approach, we have also self-consistently
predicted the halo population inside of voids.  In agreement with
naive expectations, small haloes should be less underdense than
massive haloes.  But the discrepancy is by no means large -- typically
smaller than a factor of two over three decades in halo mass.
Including satellite galaxies in this calculation (which we have not
done) will decrease the difference.  Thus, the predicted steepening of
the mass function in low-density environments is only modest and is
not ruled out by existing observations of the galaxy luminosity
function \citep{croton05,hoyle05}.  We also emphasize that these
observations determined the environmental density on relatively small
scales ($\sim 8 h^{-1} \Mpc$), where random fluctuations in the galaxy
field may mimic true underdensities.  If so, differences between mean
and low-density environments will be further washed out.  Moreover, on
such scales, finite-size effects become important in setting the
characteristic luminosity $L_\star$.  The test can be sharpened by
focusing on the luminosity function within large, easily identified
voids -- although, even there, separating void and wall galaxies may
be difficult.  Thus, we find no compelling reason to believe that galaxy
formation differs inside and outside of voids -- although we have no
evidence against such a possibility, either.

Statistical properties of the galaxy distribution can also be used to
test our predictions, especially for small voids that cannot be
identified unambiguously.  The void probability function quantifies
the probability that a sphere of a given size contains no galaxies
\citep{white79}, and the underdensity probability function quantifies
the probability that a sphere is more underdense than some specified
value \citep{vogeley89}.  These measures are free from bias in the
void selection process, but they are harder to predict directly from
the excursion set formalism and are less useful in identifying
galaxies that reside in voids.  Our model is in qualitative agreement
with the observed trends in these statistics \citep{hoyle04,conroy05},
but more work is needed to connect our formalism to them.


\appendix

\section{The First Crossing Distribution for Two Barriers}
\label{twolin}

\citet{sheth04} showed how to derive equation (\ref{eq:nvm}) using the
Laplace transform.  Here we present an alternate derivation (similar
to \citealt{mcquinn05}) and generalize it to the case of two linear
barriers with equal slope.  We begin by considering two
constant barriers $\dL_p=B_p$ and $\dL_v=B_v$.  The distribution of
trajectories $Q(\dL,S)$, where $S=\sigma^2$, obeys the diffusion
equation
\begin{equation}
\frac{\partial Q}{\partial S} = \frac{1}{2} \, \frac{\partial^2
Q}{\partial (\dL)^2}.
\label{eq:diff}
\end{equation}
The appropriate boundary conditions are $Q(B_p,S)=Q(B_v,S)=0$ and
$Q(\dL,0)=\delta_D(\dL)$, where $\delta_D$ is the Dirac delta
function.  We first make the simple transformation $y=\dL-B_v$.  We
then assume $Q(y,S) = f(S) g(y)$; equation (\ref{eq:diff}) becomes
\begin{equation}
\frac{1}{f} \, \frac{\deriv f}{\deriv S} = \frac{1}{2g} \,
\frac{\deriv^2 g}{\deriv y^2} = - \alpha^2,
\label{eq:sep}
\end{equation}
where $\alpha$ is a constant.  The general solutions are $f(S) =
\exp(-\alpha^2 S)$ and $g(y) = \exp( - \sqrt{2} \alpha y)$.  The
boundary conditions $g(0)=g(B_p-B_v)=0$ select the sine function for
the latter and force its argument to take discrete values; thus
\begin{equation}
Q(y,S) = \sum_{n=1}^\infty a_n \, \sin \left( \frac{n \pi y}{B_p-B_v}
\right) \, \exp \left[ - \frac{n^2 \pi^2 S}{2 (B_p - B_v)^2} \right].
\label{eq:Q1}
\end{equation}
We fix the constants $a_n$ by matching to the function
$Q(y,0)=\delta_D(y+B_v)$.  Thus
\begin{eqnarray}
Q(y,S) & = & \sum_{n=1}^\infty \frac{2 \, \sin (n \pi D)}{B_p-B_v} \,  \sin
\left( \frac{n \pi y}{B_p-B_v} \right) \nonumber \\ 
& & \times \exp \left[ - \frac{n^2
\pi^2 S}{2 (B_p - B_v)^2} \right],
\label{eq:Q2}
\end{eqnarray}
where $D \equiv -B_v/(B_p-B_v)$.  The total rate at which trajectories
disappear from the permitted region is
\begin{equation}
- \frac{\partial}{\partial S} \int_0^{B_p-B_v} \deriv y \, Q(y,S) =
  -\frac{1}{2} \left[ \frac{\partial Q}{\partial y}
  \right]_0^{B_p-B_v},
\label{eq:flow}
\end{equation}
where we have used equation (\ref{eq:diff}).  We interpret the two
terms on the right hand side as the rate at which trajectories flow
across the two barriers.  Thus the first-crossing distribution for the
void barrier is
\begin{eqnarray}
F_v(S) & = & \frac{1}{2} \, \left. \frac{\partial Q}{\partial
y} \right|_{y=0} \nonumber \\ & = & \sum_{n=1}^\infty \frac{(n \pi
D)^2}{B_v^2} \, \frac{\sin (n \pi D)}{n \pi} \, \exp \left[ - \frac{n^2
\pi^2 D^2}{2 B_v^2/S} \right],
\label{eq:fc-con}
\end{eqnarray}
which yields equation (\ref{eq:nvm}) after converting to mass units.

We now consider the case of two absorbing barriers linear in $S$,
$\dL_p=B_p + B_1 S$ and $\dL_v = B_v + B_1 S$.  We make the coordinate
transformation $u=B_1(\dL-B_1 S) - B_1 B_v$; then the boundary
conditions on $u$ become $Q(0,S)=0$ and $Q(u_1,S)=0$, where $u_1
\equiv B_1(B_p-B_v)$.  The boundary condition on $S$ becomes
\begin{equation}
Q(u,0) = \delta_D(\dL) = B_1 \, \delta_D(u+B_1 B_v) \, \exp(-u-B_1 B_v).
\label{eq:bcS}
\end{equation}
The diffusion equation (\ref{eq:diff}) becomes
\begin{equation}
\frac{\partial Q}{\partial S} = \frac{B_1^2}{2} \left(
\frac{\partial^2 Q}{\partial u^2} + 2 \frac{\partial Q}{\partial u}
\right).
\label{eq:diffu}
\end{equation}
We again assume a separable solution $Q(u,S) = f(S) g(u)$.  The
general solutions are exponentials, and the boundary conditions on $g$
select the sine function with discrete arguments.  Thus the solution is
again a Fourier sine series in $u$.  The Fourier coefficients $a_n$
can be fixed by matching the solution to equation (\ref{eq:bcS}).  We
then find that
\begin{eqnarray}
Q(u,S) & = & e^{-u} \, \sum_{n=1}^\infty \frac{2 \sin (n \pi D)}{B_p-B_v} \, 
\sin \left[ \frac{n \pi u}{B_1(B_p-B_v)} \right] \, 
\nonumber \\ & & \times \exp \left[ - B_1 B_v - 
\frac{B_1^2 S}{2} - \frac{n^2 \pi^2 S}{2 (B_p - B_v)^2} \right];
\label{eq:Ql1}
\end{eqnarray}
note the similarity to equation (\ref{eq:Q2}).  The rate at
which trajectories cross the barriers is
\begin{equation}
- \frac{\partial}{\partial S} \int_0^{u_1} \deriv \dL Q(\dL,S) =
  -\frac{B_1}{2} \left[ \frac{\partial Q}{\partial u} \right]_0^{u_1}
  - B_1 [Q]_0^{u_1}.
\label{eq:flowlin}
\end{equation}
But $Q(0,S)=Q(u_1,S)=0$; thus we can again identify the first-crossing
distribution of the linear void barrier $F_v^{\ell}$ with the $u=0$
component of the first part, so
\begin{eqnarray}
F_v^{\ell}(S) & = & \frac{B_1}{2} \left. \frac{\partial
Q}{\partial u} \right|_0 \nonumber \\
& = & F_v(S) \, \exp(-B_1 B_v - B_1^2 S/2),
\label{eq:fc-lin}
\end{eqnarray}
from which the mass function follows directly.  Note that this
expression differs from equation (C10) of \citet{sheth04}.  Those
authors tried to derive the same distribution through Laplace transforms.
Their solution required the first-crossing distribution of $\dL_p$ and
$\dL_v$ to be identical (their equation C6); this is not the
case if the barriers have equal slope $B_1$, because then they are not
symmetric about the $\dL=0$ axis.  Instead their solution
applies to a case in which the slopes of the two barriers have opposite signs.

\end{document}